\documentclass{aa}
\usepackage{graphicx}

\def\msol{\hbox{\kern 0.20em $M_\odot$}}
\def\lsol{\hbox{\kern 0.20em $L_\odot$}}
\def\rsol{\hbox{\kern 0.20em $R_\odot$}}
\def\sr{\hbox{\kern 0.20em sr}}
\def\srmu{\hbox{\kern 0.20em sr$^{-1}$}}
 
\def\g{\hbox{\kern 0.20em g}}
\def\gmu{\hbox{\kern 0.20em g$^{-1}$}}
\def\kg{\hbox{\kern 0.20em kg}}
\def\pc{\hbox{\kern 0.20em pc}}
 
\def\mum{\hbox{\kern 0.20em $\mu$m}}
\def\mumd{\hbox{\kern 0.20em $\mu$m$^{-2}$}}
\def\cm{\hbox{\kern 0.20em cm}}
\def\m{\hbox{\kern 0.20em m}}
\def\km{\hbox{\kern 0.20em km}}
\def\nm{\hbox{\kern 0.20em nm}}
 
\def\s{\hbox{\kern 0.20em s}}
\def\h{\hbox{\kern 0.20em h}}
\def\sec{\hbox{\kern 0.20em sec}}
\def\min{\hbox {\kern 0.20em min}}
\def\smu{\hbox{\kern 0.20em s$^{-1}$}}
\def\smd{\hbox{\kern 0.20em s$^{-2}$}}
\def\an{\hbox{\kern 0.20em an}}
\def\anmu{\hbox{\kern 0.20em an$^{-1}$}}
\def\deg{\hbox{\kern 0.20em $^{\rm o}$}}
\def\yr{\hbox{\kern 0.20em yr}}
\def\yrmu{\hbox{\kern 0.20em yr$^{-1}$}}
\def\Myr{\hbox{\kern 0.20em Myr}}
\def\Mymu{\hbox{\kern 0.20em Myr$^{-1}$}}
 
\def\pcmu{\hbox{\kern 0.20em pc$^{-1}$}}
\def\pcmd{\hbox{\kern 0.20em pc$^{-2}$}}
\def\pcmt{\hbox{\kern 0.20em pc$^{-3}$}}
\def\kms{\hbox{\kern 0.20em km\kern 0.20em s$^{-1}$}}
\def\kmpd{\hbox{\kern 0.20em km$^{2}$}}
\def\kpc{\hbox{\kern 0.20em kpc}}
\def\cms{\hbox{\kern 0.20em cm\kern 0.20em s$^{-1}$}}
\def\erg{\hbox{\kern 0.20em erg}}
\def\ergs{\hbox{\kern 0.20em erg}}
\def\cmpd{\hbox{\kern 0.20em cm$^2$}}
\def\cmmd{\hbox{\kern 0.20em cm$^{-2}$}}
\def\cmms{\hbox{\kern 0.20em cm$^{-6}$}}
\def\cmpt{\hbox{\kern 0.20em cm$^3$}}
\def\cmmt{\hbox{\kern 0.20em cm$^{-3}$}}
\def\mpd{\hbox{\kern 0.20em m$^2$}}
\def\mmd{\hbox{\kern 0.20em m$^{-2}$}}
\def\mpt{\hbox{\kern 0.20em m$^3$}}
\def\mmt{\hbox{\kern 0.20em m$^{-3}$}}
\def\mum{\hbox{\kern 0.20em {\rm $\mu$m}}}
\def\mujy{\hbox{\kern 0.20em $\mu$Jy}}
\def\mjy{\hbox{\kern 0.20em mJy}}
\def\Mj{\hbox{\kern 0.20em MJy}}
\def\jy{\hbox{\kern 0.20em Jy}}
\def\ghz{\hbox{\kern 0.20em GHz}}

\def\G{\hbox{\kern 0.20em G}}
\def\K{\hbox{\kern 0.20em K}}

\def\h13cop{\hbox{H$^{13}$CO$^{+}$}}

\def\S+{\hbox{S{\small II}}}
\def\H+{\hbox{H{\small II}}}

\begin{document}

\title{Disks around Hot Stars in the Trifid Nebula\\
}

\subtitle{}

\author{
B.~Lefloch
  \inst{1}
\and J.~Cernicharo
  \inst{2}
\and D.~Cesarsky
  \inst{3}
\and K.~Demyk
  \inst{4}
\and M.A.~Miville-Desch\^enes
  \inst{5}
\and L.F.~Rodriguez
   \inst{5}
  }

\offprints{lefloch@obs.ujf-grenoble.fr}

\institute{
  Laboratoire d'Astrophysique de l'
  Observatoire de Grenoble,
  BP 53, 38041 Grenoble Cedex,
  France
  \and 
  Consejo Superior de Investigaciones Cient\'{\i}ficas,
  Instituto de Estructura de la Materia, 
  Serrano 123, E-28006 Madrid
  \and
  Max-Planck Institut f\"{u}r Extraterrestrische Physik, 
  85741 Garching,  Germany
  \and
  Institut d'Astrophysique Spatiale, 
  B\^{a}t. 121, Universit\'e Paris XI, 
  91450 Orsay Cedex 
  France
  \and
  Ecole Normale Sup\'erieure, 
  LRA, 
  24 rue Lhomond, 75231, Paris, France
  \and
  Instituto de Astronom\'{\i}a, UNAM, 
  Campus Morelia, A.P. 3-72, 
  Morelia,   Mich. 58089,
  M\'exico
  }

\date{Received 08 December 2000; accepted 09 January 2001}

\abstract{
We report on mid-IR observations of the central region in the Trifid
nebula, carried out with ISOCAM in several broad-band infrared
filters and in the low resolution spectroscopic mode provided by the
circular variable filter. 
Analysis of the emission indicates the presence of 
a hot dust component (500 to 1000~K) and a warm dust component at lower
temperatures ($\sim 150-200$~K) around several members of the cluster
exciting the H{\small II} region, and other stars undetected at optical 
wavelengths. Complementary VLA observations suggest that the mid-IR emission
could arise from a dust cocoon or a circumstellar disk, evaporated
under the ionization of the central source and the exciting star of the 
nebula. In several sources the $\rm 9.7\mu m$ silicate band is seen in 
emission. One young stellar source shows indications of 
crystalline silicates in the circumstellar dust. 
}

\maketitle
 
\section{Introduction}

The Trifid nebula is a nice example of a small H{\small II} region
in an early stage of evolution. It is  located at a distance of 
$1.68\kpc$ (Lynds et al. 1985) and has an estimated dynamical age 
of $0.3-0.4\Myr$ (Lefloch \& Cernicharo 2000). Because of its youth and 
its small size, it provides
us  with a comprehensive picture of the early stages of massive
star-forming regions like Orion. Indeed, previous observations at 
millimeter wavelengths have shown that the protostellar cores 
surrounding the H{\small II} region have physical 
properties similar to the cores observed in Orion. These protostellar  
cores show good evidence that their birth was induced in the expansion 
of the  H{\small II} region (Lefloch \& Cernicharo 2000).  

The Trifid is  excited by the star HD~164492A, classified as O7~V by 
Lynds et al. (1985). Five other stars (HD~164492B-F)
were identified within $20\arcsec$ ($0.16\pc$) of HD~164492A, 
suggesting that the formation of a cluster accompanied the birth of the 
ionizing star. The spectral types
determined for some of the sources indicate they are of intermediate mass
(Kohoutek, 1997). 
Here we report on mid-infrared ISOCAM observations of the central region 
in the H{\small II} region. 
These observations bring more insight of the star formation which accompanied
the birth of the ionizing star.  In particular, they reveal the lower 
mass star formation, embedded in the low-density ionized gas of the 
H{\small II} region. 
We find that many sources are surrounded by dense material, 
possibly in the form of disks photoevaporated by the  ionizing radiation 
of the Trifid, similar to those found close to the Trapezium stars in Orion.

\section{Observations and Results}

\subsection{The Mid-infrared Emission}

Several images of the Trifid were obtained with ISOCAM onboard ISO
(Cesarsky et al. 1996) in the LW4 filter centered on the PAH band at 
$6.2\rm \mum$ ($\Delta\lambda= 1.0\rm \mum$) at  $6\arcsec$ resolution, 
in the  LW7 filter centered on the 
silicate band at $10\rm \mum$ ($\Delta\lambda= 3\rm \mum $) at  $6\arcsec$
resolution and in the LW10 filter
($\lambda= 11.5\rm \mum$,  $\Delta\lambda= 7\rm \mum$) at $3\arcsec$ 
resolution. The final maps are shown in Fig.~1.
Several maps of the Trifid were obtained with the Circular Variable Filter 
(CVF) between 5 and $17\rm \mu m$, with a spectral resolution 
$\lambda/\Delta\lambda\simeq 30$. 
 Each map has a size of $32\times32$ pixels
with an angular size of $3\arcsec$ per pixel. 

Superposed to the extended emission from the dust lanes, we detect in the 
central LW4 image 
several point-like sources (IRS~1 to IRS~5) near the stellar cluster
(see Fig.~1d) and two more remote sources (IRS~$6-7$), apparent on 
Fig.~1c-e.
The pointlike sources appear with the highest contrast in the LW4 filter; 
by comparison, IRS~4 and IRS~5 are hardly visible in the LW7 and LW10
band. Except for the bright sources IRS~$1-2$ and IRS~6, the emission of 
the individual sources in the LW7 filter can hardly be separated from 
the extended emission of the dust lanes. 

The maximum of emission (source IRS~$1-2$) peaks approximately $15\arcsec$
Southwest of HD~164492A and coincides with the members of the 
central cluster HD~164492C-D. They are optically visible (Fig.~1a) 
and were classified as B6~V (Gahm et al. 1983) and Be LkH$\alpha 123$ 
(Herbig, 1957) respectively. 
In the LW4 filter, the flux distribution is
marginally resolved and indicates a size of $\sim 3\arcsec$ for the 
emitting region. The angular resolution of the infrared observations
does not allow to distingush the respective contributions of components C
and D. However, a gaussian fit to the CVF data gives a size 
of $2.5\arcsec$ for the emitting region. 
This size is very close to the projected distance between 
components C and D ($\sim \rm 4000~AU$, Sect. 2.3) 
and suggests that both sources are contributing to the 
infrared flux; we identify IRS~1 (IRS~2) with component C (D) of the
stellar cluster. All the other infrared sources are unresolved in our 
images. 

\begin{figure*}
  \begin{center}
    \leavevmode
  \end{center}
\caption{
{\bf a)} Optical image
of the $\rm H\alpha$ emission in the Trifid nebula taken with the
IAC80 telescope at the Observatorio del Teide, Tenerife, Spain (see also 
Cernicharo et al. 1998). Coordinates
are in arcsec offsets with respect to the exciting star of the nebula
HD~164492A : $\alpha_{2000}$= $\rm 18^h 02^m 23.55^s$,
$\delta_{2000}= -23^{\circ}01\arcmin 51.0\arcsec$
{\bf b)} Infrared emission in the $8-15\rm \mum$ band (LW10).
The black star marks the position of HD~164492A.
{\bf c)-d) } Infrared emission in the $5.5-6.5\rm \mum$ band (LW4) and 
$8.5-11.5\rm \mum$ band (LW7) respectively.  
{\bf e)} Magnified view of the $\rm H\alpha$ emission.
We have superposed (contours) the 3.6cm free-free
emission observed at the VLA. The contours range from 0.2 to $1\mjy$ 
by $0.2\mjy\, \rm beam^{-1}$. {\bf f)} Spectral emission 
 observed with the CVF towards IRS~1-2
and fit to the continuum. 
{\bf g)}~Silicate emission band observed towards IRS~1-2 after 
subtracting the continuum (thick) and fit to the emission (thin) 
assuming a mix of pure forsterite with amorphous pyroxenes with inclusions 
of FeO. 
}
\end{figure*}

\subsection{The physical properties of the dust}

The spectra of IRS~1-2, IRS~3 and IRS~6 between 5 and $17\rm \mum$ were
obtained from the CVF data after subtracting the emission of a 
nearby 3 x 3 pixel$^2$ area used as a reference.
The observed emission was fitted with a gaussian function convolved 
with the Point Spread Function of the instrument. 
The IRS~1-2 spectrum is shown in Fig.~1f. It does not exhibit 
any narrow emission, apart from a weak Ne{\small II} line at $12.7\rm \mum$. 
Remarkably, the 
silicate band at $9.7\rm \mum$ is detected in {\em emission}  
whereas it is seen in {\em absorption} in the
dust lanes (not shown here). 

We model the dust emission as a black-body law modified by a dust 
opacity $\tau_{\nu} \propto \nu$.
The $5-17\rm \mum$ continuum emission is satisfactorily fitted by two 
components
at temperatures $T\sim 500$~K and $T\sim 200$~K respectively (Fig.~1f). 
We adopted $0.2\arcsec$ for the size of the ``cold'' component, as determined 
from the VLA observations (see below). The lack of constraints on the size 
of the hot component makes the amount of material difficult to quantify
(especially the hot component). We derive typical hydrogen column densities
$\rm N(H)\sim 10^{21}\cmmd$ for the component at 200~K. 
The IRS~3 and IRS~6 spectra taken with the CVF display properties similar to 
IRS~1-2 (not shown). The continuum emission can be accounted for by the 
same simple model~: one hot layer at $\sim 700-1000$~K and a warm
component at $\sim 150-300$~K.

The shape of the silicate band around $9.5\rm \mum$ (Fig.~1g) 
can help constrain the composition of the dust. 
The blue wing of the band (8-10 $\mu$m) 
is well reproduced with amorphous pyroxene grains 
containing FeO inclusions (we assume spherical grains of $0.1\rm \mum$ size 
and the optical constants are taken from Henning et al. 1999). 
Comparison with 
laboratory transmission spectra of submicronic grains obtained at the IAS
suggests that the broad shoulder longward of $11.5\rm \mum$ is probably due to
crystalline silicates; it is well fitted with pure forsterite grains 
(Fig.~1g). We note that other crystalline silicates, such as pyroxenes,  
could also be present and account for the red shoulder. However, 
observations of other bands longward of $15 \rm \mum$,
like the $33.6\rm \mum$ forsterite band or the $40.5\rm \mum$ pyroxene 
band are necessary to confirm our identification and search for other disk 
components, like crystalline pyroxenes.

\subsection{The ionized material}

We observed the continuum radiation of the central region 
at 3.6cm using the Very Large Array in its highest angular resolution 
configuration in 1998 March 13. 
The field was centered at $\rm \alpha= 18^h02^m27.223^s$ $\delta= -23\deg 
03\arcmin 15.91\arcsec$ (Eq.~2000), so that 
all the infrared sources but IRS6 lie within the field. 
We used 1328+307 as absolute amplitude calibrator and 1748-253 as the
phase calibrator. A bootstrapped flux density of 0.271$\pm$0.001 Jy was
obtained for 1748-253. The observations were made in both circular
polarizations with an effective bandwidth of 100 MHz.  
The rms noise in the map is $20 \mu$Jy beam$^{-1}$; the beam size (HPFW) is 
$0.38\arcsec \times 0.19\arcsec$ and the position angle is $11^\circ$. 
The map reveals the presence of two radio
sources that coincide respectively with components C and D of the stellar 
cluster. We do not detect any emission from any other 
ISOCAM source, nor component B of the cluster, for which 
Yusef-Zadeh et al. (2000)  report a 3.6~cm flux of 0.66 mJy, much higher 
than our $5-\sigma$ limit of $0.1\mjy$. 
It appears unlikely that this emission arises from a circumstellar 
disk, as suggested by these authors;  
we propose that component B  
is a time variable gyrosynchrotron emitter, that may exhibit circular
polarization in high-sensitivity radio-observations. 

The emission from components C and D is compact and forms a close 
double, separated by 
$2.4\arcsec$. The flux densities at 3.6cm were $1.64\pm 0.02\mjy$ 
(for C) and $1.48\pm 0.02\mjy$ (for D). Both sources appear 
unresolved with an angular size smaller than $\sim 0.2\arcsec$. 
No linear or circular polarization was found in the sources
to an upper limit of $\sim$4\% for the degree of polarization. 
The lack of polarization is consistent with a 
free-free interpretation for the observed emission. To obtain the spectral 
indices of both objets we undertook additional 
VLA observations at 2-cm in the C configuration during 1998 November 21.
We used 0134+329 as absolute amplitude calibrator and 1741-312 as the
phase calibrator, with a bootstrapped flux density of 0.574$\pm$0.003 Jy.
The flux densities at 2-cm were 1.66$\pm$0.14 mJy 
(for C) and 1.30$\pm$0.14 mJy (for D), that combined with the 
3.6-cm measurements imply spectral indices of 0.0$\pm$0.2 and 
-0.2$\pm$0.2, for C and D, respectively. These flat spectral indices 
in the centimeter range are characteristic of optically thin free-free
emission. 
The measured fluxes  are in rough agreement with similar observations 
at a lower angular resolution ($\sim 0.5\arcsec$) by 
Yusef-Zadeh et al. (2000).
From the observed emissivities and adopting a transverse size of 
$0.2\arcsec$ for the emitting region, we estimate a density of 
$\approx 10^5\cmmt$ and a total mass of $6.0\times 10^{-6}\msol$ 
for the ionized gas around C and D. 

\section{Discussion}

The color excesses measured by Kohoutek et al. (1997) 
towards the central stellar cluster ($\sim 0.3-0.4$)
indicate very low hydrogen column densities 
$\rm N(H)= 1.9-2.6\times 20^{21}\cmmd$. Hence, there 
is no large-scale gas reservoir which could slow down the expansion 
of the ionized gas detected around components C and D. The ionized gas 
must expand 
on the sonic timescale $\tau_i= R_i/c_i \sim 80\yr$,  with $R_i= 0.1\arcsec$ 
as the radius of the region observed and $c_i= 10\kms$. 
This timescale is so short compared to the age of the
nebula that the free-free emission has to be sustained through the 
ionization of some ``fresh'' neutral material.
The mid-infrared data provide direct evidence for some neutral 
material around components C and D. 
Therefore, we propose that the emission observed at the VLA and in the
infrared originates from a reservoir of dense circumstellar material 
exposed to the ionizing radiation of HD~164492A, 
similar to the photo-evaporated disks detected in Orion 
(Churchwell et al. 1987; O'Dell, Wen \& Hu 1993). 

Following the same approach as Churchwell et al. (1987), we estimate an 
ionizing flux $J_i= 1.6\times 10^{11}\cmmd\smu$ at the ionization front and 
a mass-loss rate $\rm \dot{M}\simeq 4\times 10^{-8}\msol\yrmu$. 
It implies the mass of warm material detected in the infrared 
($\sim 2-3\times 10^{-5}\msol$) would have completely evaporated on a 
timescale of $400\yr$, whereas the nebula is $\sim 0.3\Myr$ old.
The total circumstellar mass is probably much higher than the mass 
of material detected and the bulk of disk material corresponds 
to a colder component which emits at longer wavelengths.
We also note that the mass of the photo-evaporated disks 
discovered in the Orion nebula are 100 times larger than 
the masses estimated here (St\"{o}rzer and Hollenbach, 1999). 
The similarity between the mass of the ionized gas and the mid-infrared 
material suggests that we are detecting the emission of the 
photon-dominated region at the surface of the circumstellar disks. 
Interestingly, the column density of warm gas derived from the CVF data 
is in good agreement with the model of Johnstone, Hollenbach \& Bally (1998)
for the warm PDR gas at the surface of a photo-evaporated disk. 

The only two infrared sources detected at the VLA are classified as 
massive or intermediate-mass objects; this suggests  
that all the infrared sources detected with ISOCAM have also 
low- or intermediate masses. This would naturally explain why 
neither IRS~3 nor IRS~4 were detected at the VLA, though 
at similar apparent distances to the ionizing
source. However, we cannot exclude that we are misled by
a projection effect and that the true distance to the 
exciting star is much larger. 
Apart from IRS~1-2, IRS~3  and IRS~6 exhibit the silicate band at 
$9.7\rm \mum$ 
in emission although both lay behind the Western dust lane of the 
nebula. It means that the dust temperature of the circumstellar reservoir 
must be at least $\sim  200$~K.  
Since IRS~3 is very close to HD~164492A and is also detected in the 
optical $\rm H\alpha$ line, one cannot exclude that the circumstellar 
dust is externally heated by the exciting star. 
Interestingly, this is not the case for other sources like IRS~5. This is
probably because these sources are still embedded in their cocoon,
and the outer heating is not sufficient to compete with the inner continuum
source.   

Only very few examples of crystalline silicates in protostellar disks
have been reported so far (Malfait et al. 1998, 1999). The mechanism
responsible for the crystallization of silicates is unclear (Molster et
al. 1999). Our observations in the Trifid nebula show that many young stellar 
sources are removing their parental cocoon, leaving their dense
circumstellar disk exposed to the ionizing radiation of the exciting
stars. In some sources we find a dust component at 
temperatures comparable to the glass temperature of silicates ($\sim 10^3$~K).
We speculate that dust annealing in the PDR of the photoionized disks could
contribute and maybe account for the crystalline component detected
in IRS~1-2 (see e.g. Gail, 1998). 
Complementary observations of IRS~1-2 with upcoming infrared instruments like 
SIRTF are necessary to confirm the presence of crystalline silicates, 
characterize their physical properties, and to constrain the possibility 
of such a scenario. 

\begin{center} {\it Acknowledgements} \end{center}
We acknowledge Spanish DGES for this supporting 
research under grants PB96-0883 and ESP98-1351E. LFR is grateful
to CONACyT, Mexico, for its support
This research made use of SIMBAD.

\end{document}